\documentclass[12pt]{article}
\begin{document}

\begin{flushright}
\hfill CAMS/98-05\\ \hfill hep-th/9809148\\
\end{flushright}
\vspace{1cm}

\begin{center}
\baselineskip=16pt {\Large \textbf{Chiral Anomalies in the
Spectral Action \footnote{To appear in the book "Quantum Groups
and Fundamental Physical Interactions", Editor: D. Kastler,
Publisher: Nova Science Publishing Company. }}}

\vskip 1 cm \vskip 2cm \textbf{Ali H.
Chamseddine\footnote{\emph{Email chams@aub.edu.lb ,
chams@itp.phys.ethz.ch}}}
\\[0pt]
\vskip 0.5cm \emph{Center for Advanced Mathematical Sciences\footnote{\emph{%
Permanent address.}} \\[0pt] American University of Beirut\\[0pt]
Beirut, Lebanon\\[0pt] \vskip 0.5cm and\\[0pt] Institut f\"ur
Theoretische Physik\\[0pt] ETH H\"onggerberg \\[0pt] CH-8093
Z\"urich, Switzerland}\\[0pt]
\end{center}

\vskip 1cm

\begin{abstract}
The definition of the spectral action involves the trace operator
over states in the physical Hilbert space. We show that in the
presence of chiral fermions there are consistency conditions on
the fermionic representations. These conditions are identical to
the conditions for absence of gauge and gravitational anomalies
obtained in the path integral formalism.
\end{abstract}

\bigskip \bigskip \newpage At present we have a great deal of information
about the particle physics spectrum at low-energies. Using the
tools of noncommutative geometry it is now possible to use the
particle physics spectrum to study the geometry of space-time
\cite {Connes,Cobook,CoLo,Kastler,CFF} . This program has been
advocated in \cite {ACAC} where it is shown that a real spectral
triple corresponding to a product of a continuous space by a
discrete space serves as a good starting point. In \cite {ACAC} a
new principle is proposed, the spectral action principle. This
states that the action for the dynamical degrees of freedom
contained in the metric is given by some function of the perturbed
Dirac operator. The results do not depend crucially on the form of
the function used in the spectral action, but mainly on the
spectrum of the Dirac operator. In other words, the spectrum
contains all the information about geometric invariants and should
help in indicating the relevant geometry of space-time.

It is well known that fermions in the standard model are chiral,
and that masses are generated when the Higgs field acquires a vev
so that the Higgs-fermi-fermi interaction produces a fermionic
mass term. The trace operator is defined by summing over
eigenstates of the Dirac operator. In the case of chiral fermions,
the eigenvalue problem for the Dirac operator cannot be defined
without adding fermions with opposite chirality. This is the
familiar problem encountered in the path integral formalism, and
there one assumes the spinors and their conjugates to be
independent. The trace operator, unlike the fermionic kinetic
energy, is not invariant under chiral rotations. The invariance is
spoiled by the appearance of a phase factor which must be made to
vanish for the action to be consistent.

First, I shall briefly review the ingredients that enter the
derivation of the spectral action for the standard model, and then
show how this action transforms under chiral rotations. The
subtleties involved in working with Euclidean signature, as well
as with Minkowski signature, are also discussed \cite {L,BGS}.

A noncommutative space is defined by a spectral triple $\left( \mathcal{A},%
\mathcal{H},D\right) \,$ where $\mathcal{A}$ is an algebra of operators, $%
\mathcal{H}$ a Hilbert space of states, $D$  an unbounded operator in $%
\mathcal{H}$. The space is supplied with a real structure \cite
{Co95} $J$ satisfying
\begin{equation}
J^{2}=\epsilon ,\qquad JD=\epsilon ^{^{\prime }}DJ,\qquad J\gamma =\epsilon
^{^{\prime \prime }}\gamma J;
\end{equation}
\[
\epsilon ,\epsilon ^{^{\prime }},\epsilon ^{^{\prime \prime }}\in \left\{
-1,1\right\} ,
\]
where the value of $\epsilon ,\epsilon ^{^{\prime }},\epsilon ^{^{\prime
\prime }}$ is determined by $n$ modulo $8$. The algebra $\mathcal{A}$ is
taken to be
\begin{equation}
\mathcal{A}=C^{\infty }(M)\otimes \mathcal{A}_{F},
\end{equation}
where the algebra $\mathcal{A}_{F}$ is finite dimensional
\begin{equation}
\mathcal{A}_{F}=C\oplus H\oplus M_{3}(C),
\end{equation}
and $H$ is the algebra of quaternions
\begin{equation}
H=\left\{ \left(
\begin{array}{ll}
\alpha & \beta \\
-\overline{\beta } & \overline{\alpha }
\end{array}
\right) ;\alpha ,\beta \in C\right\} .
\end{equation}
Similarly,
\begin{eqnarray}
\mathcal{H} &=&L^{2}(M,S)\otimes \mathcal{H}_{F}, \\ D
&=&D_{M}\otimes 1+\gamma _{5}\otimes D_{F},
\end{eqnarray}
where $\left( \mathcal{A}_{F},D_{F}\right) $ is a spectral geometry on $%
\mathcal{A}_{F},$ while $L^{2}(M,S)$ is the Hilbert space of
$L^{2}$ spinors and $D_{M}$ is the Dirac operator of the
Levi-Civita connection. The list of elementary fermions provide a
natural candidate for $\mathcal{H}_{F}.$ One lets
$\mathcal{H}_{F}$ be the Hilbert space with basis labeled by
elementary leptons and quarks. The $Z_{2}$ grading $\gamma _{F}$
is given by +1 for left handed particles and -1 for right handed
ones. The involution $J$ is such that $Jf=\overline{f}$ for any
$f$ in the basis. One has $J^{2}=1$ and $J\gamma =\gamma J.$ The
fermions are represented by
\begin{equation}
\Psi =\left(
\begin{array}{l}
f \\
\overline{f^{\prime }}
\end{array}
\right) ,
\end{equation}
where$\overline{f}\in \overline{\mathcal{H}}$,
$\overline{\mathcal{H}}$ being the conjugate Hilbert space. In
this representation the action of $J$ is given by
\begin{equation}
J\Psi =\left(
\begin{array}{l}
f^{\prime } \\
\overline{f}
\end{array}
\right) ,
\end{equation}
while $D_{F}$ is given by
\begin{equation}
D_{F}=\left(
\begin{array}{ll}
Y & 0 \\
0 & \overline{Y}
\end{array}
\right) ,
\end{equation}
where $Y$ is the Yukawa coupling matrix. The group
$Aut(\mathcal{A})$ of automorphisms of the involutive algebra
$\mathcal{A}$ plays the role of
diffeomorphisms of the noncommutative geometry. It has a normal subgroup $%
Int(\mathcal{A)}$ where an automorphism $\alpha $ is inner iff there exists
a unitary operator $u\in \mathcal{A},(uu^{*}=u^{*}u=1)$ such that
\begin{equation}
\alpha (a)=uau^{*},\qquad \forall a\in \mathcal{A} .
\end{equation}
The group $Aut(\mathcal{A})$ of diffeomorphisms falls in equivalence classes
under the normal subgroup of inner automorphisms. In the same way the space
of metrics has a natural foliation into equivalence classes. The internal
fluctuations of a given metric are given by the formula
\begin{eqnarray}
D &=&D_{0}+A+JAJ^{-1}, \\
A &=&\sum\limits_{i}a_{i}\left[ D_{0},b_{i}\right] ,
\end{eqnarray}
where $a_{i},b_{i}\in \mathcal{A}$ and $A=A^{*}$ . Starting with
$\left( \mathcal{A},\mathcal{H},D_{0}\right) $ one leaves the
representation of $\mathcal{A}$ in $\mathcal{H}$ and perturbs the operator $%
D_{0}.$

The spectral action that reproduces the standard model Lagrangian takes the
very simple form
\begin{equation}
\left( \Psi \left| D\right| \Psi \right) +TrF\left(
\frac{D}{m}\right) .
\end{equation}
where $m$ is a cut-off scale, needed to make the argument of the
function $F$ dimensionless. It is shown in \cite {ACAC} that the
fermionic action reproduces the fermionic kinetic energies as well
as their gauge and Higgs interactions. The bosonic action
contains, to lowest orders, a cosmological constant, the curvature
scalar, Higgs mass term as well as Weyl gravity coupled
conformally to the Higgs and gauge fields. The dependence on the
function $F$, to the lowest orders indicated, enters only through
three parameters, $f_o$, $f_2$ and $f_4$ which multiply the
cosmological constant, the curvature and Higgs terms and the
conformal terms respectively, where
\begin{eqnarray}
f_{0} &=&\int\limits_{0}^{\infty }F(u)u du , \\ f_{2}
&=&\int\limits_{0}^{\infty }F(u)du ,\\ f_{2(n+2)} &=&\left(
-1\right) ^{n}F^{(n)}(0).
\end{eqnarray}
Higher order terms are suppressed by $\frac{1}{m}$ factors.

Fermions that appear in the physical Hilbert space in any realistic model
must be chiral. The grading operator $\gamma $ is such that
\begin{eqnarray}
\gamma \psi _{\pm } &=&\pm \psi _{\pm }  \nonumber ,\\ \gamma
D\psi _{\pm } &=&\mp D\gamma \psi _{\pm } ,
\end{eqnarray}
which implies that $D\psi _{\pm }$ have opposite chirality to $\psi _{\pm }$%
. Therefore, for chiral fermions with non-zero eigenvalues, one cannot set
the eigenvalue problem for $D$ but only for $D^{2}$. We can write
\begin{equation}
D^{2}\psi _{n\pm }=\lambda _{n}\psi _{n\pm },\qquad \lambda
_{n}=\lambda _{n}^{*} .
\end{equation}
At this point, it is important to mention that the eigenvalue
problem is solved for spaces with Euclidean signature. In this
case it is also necessary to define the conjugate fermions
$\overline{\psi _{n\pm }}$ which are taken to be independent of
$\psi _{n\pm }$. Indeed, for Euclidean signature, $\psi _{n\pm }$
and their complex conjugates have the same chirality. This makes
an expression like $\left( \psi _{n\pm }\left| D\right| \psi
_{n\pm }\right) $, which is relevant for the kinetic energy,
vanishes identically. The way out is to define the inner product
to contain $\overline{\psi _{n\pm }}D\psi _{n\pm }$ and where one
assumes that $\overline{\psi _{n\pm }}$ is independent of $\psi
_{n\pm }$ and is of opposite chirality \cite{L,BGS}. This doubling
of fermions is the same encountered in the Fujikawa path integral
treatment of chiral fermions \cite{Fuji}. It is also clear that
this doubling is not necessary in a space with Minkowski signature
as the conjugate fermions have opposite chirality. By normalising
$\left( \psi _{n\mp }\left| \psi _{n\pm }\right. \right) $ to one,
we can write $\lambda _{n}=\left( \psi _{n_{\mp }}\left|
D^{2}\right| \psi _{n\pm }\right) .$ Therefore, both in the
fermionic kinetic energy term as well as in the bosonic trace one
encounters the problem of doubling of fermions. This is again
completely parallel to the path integral formalism where doubling
is needed to define both the fermionic kinetic energy and the
fermionic measure. One cannot resist mentioning that there is one
particular case where a Majorana type term involving the fermions
without their complex conjugates could be written. This happens if
one chooses the real structure $J$ to have dimension 2 (mod 8) as
one can then impose the condition $J \Psi =\overline{\Psi} $. A
realisation of this is provided by the $\underline{16}$ spinor
representation of $SO(10)$ where one can fit all  quarks and
leptons, and in addition a right-handed neutrino. Unfortunately,
when this representation is used for the standard model, the only
Majorana interaction generated without violating the standard
model quantum numbers is for the right-handed neutrino. Therefore
in the case of  the standard model  this property of $J$ does not
solve the fermions doubling problem.

Assuming that all fermions in the spectrum have positive chirality (negative
chirality fermions could always be written as the charge conjugates of
positive chirality fermions). The bosonic spectral action in this case could
be written as
\begin{equation}
I_{b}=Tr\left( F(D^{2})\right) =\sum\limits_{n}\left( \psi
_{n-}\left| F(D^{2}\right| \psi _{n+}\right)  \label{bosact} ,
\end{equation}
where $D^{2}$ could be replaced by $D_{-}D_{+}$ when acting on $\psi _{n+}$.

Let us now consider the behavior of this action under chiral
transformations. The fermionic action is invariant under chiral rotations
(suppressing the subscript $n$),
\begin{eqnarray}
\left| \psi _{+}\right) &\rightarrow &e^{i\theta \gamma
_{5}}\left| \psi _{+}\right) ,  \nonumber \\ \left( \psi
_{+}\right| &\rightarrow &\left( \psi _{+}\right| e^{i\theta
\gamma _{5}} ,
\end{eqnarray}
which implies that
\begin{equation}
\left( \psi _{+}\left| D\right| \psi _{+}\right) \rightarrow \left( \psi
_{+}\left| e^{i\theta \gamma _{5}}De^{i\theta \gamma _{5}}\right| \psi
_{+}\right) .
\end{equation}
It is a simple matter to see that
\[
e^{i\theta \gamma _{5}}De^{i\theta \gamma _{5}}=D,
\]
implying the invariance of the fermionic action.

Under chiral rotations the bosonic action (\ref{bosact}) transforms as
\begin{eqnarray}
I_{b} &\rightarrow &Tr\left\{ \sum\limits_{n}\left( e^{-i\theta
^{a}T^{a}\gamma _{5}}\psi _{n-}\right| F(D^{2})e^{i\theta
^{a}T^{a}\gamma _{5}}\left| \psi _{n+}\right) \right\}  \nonumber
\\ &=&Tr\left\{ e^{2i\theta ^{a}T^{a}\gamma _{5}}\left( \psi
_{n-}\right| F(D^{2})\left| \psi _{n+}\right) \right\} ,
\label{bostra}
\end{eqnarray}
where $T^{a}$are matrix representations for the fermions. We now
use the heat-kernel expansion for the function $F$ \cite{Gilkey}
\begin{equation}
F\left( P\right) =\sum\limits_{n\geq 0}f_{n}e_{n}\left( P\right) ,
\end{equation}
where $e_{n}\left( P\right) $ are geometric invariants whose trace gives the
Seeley-de Witt coefficients for the operator $P=D^{2}$. Under an
infinitesimal transformation equation (\ref{bostra}) simplifies to
\begin{equation}
I_{b}\rightarrow I_{b}+\sum\limits_{n\geq 0}f_{n}Tr\,\left( 2i\theta
^{a}T^{a}\gamma _{5}e_{n}\left( P\right) \right) .
\end{equation}
Because of the presence of $\gamma _{5}$ in the trace, the first
non-vanishing gauge term comes from $e_{4}\left( P\right) $ and is of the
form
\begin{eqnarray}
&&2i\theta ^{a}Tr\,\left( \gamma _{5}\gamma ^{\mu \nu }\gamma ^{\rho \sigma
}T^{a}T^{b}T^{c}\right) G_{\mu \nu }^{b}G_{\rho \sigma }^{c}  \nonumber \\
&=&i\epsilon ^{\mu \nu \rho \sigma }\theta ^{a}G_{\mu \nu }^{b}G_{\rho
\sigma }^{c}Tr\,\left( T^{a}\left\{ T^{b},T^{c}\right\} \right) .
\end{eqnarray}
Therefore the bosonic action is non-invariant, except when the condition on
the fermionic group representations
\begin{equation}
Tr\,\left( T^{a}\left\{ T^{b},T^{c}\right\} \right) =0,
\end{equation}
is satisfied. We can relate the anomaly generating term to non-conservation
of currents. Let $\theta =\theta ^{a}\left( x\right) T^{a}$, then
\begin{equation}
\left( \psi _{+}\right| D\left| \psi _{+}\right) \rightarrow
\left( \psi _{+}\right| D\left| \psi _{+}\right) -i\theta
^{a}D_{\mu }\left( \psi _{+}\right| \gamma ^{\mu }T^{a}\left| \psi
_{+}\right) ,
\end{equation}
where we have used the identity
\[
e^{i\theta \gamma _{5}}De^{i\theta \gamma _{5}}=D+i\gamma ^{\mu
}\gamma _{5}D_{\mu }\theta ^{a}T^{a} ,
\]
and integrated by parts. Defining the fermionic current
\begin{equation}
j^{\mu a}=\left( \psi _{+}\left| \gamma ^{\mu }T^{a}\right| \psi
_{+}\right) ,
\end{equation}
we then have
\begin{equation}
D_{\mu }j^{\mu a}=-\epsilon ^{\mu \nu \rho \sigma }G_{\mu \nu }^{b}G_{\rho
\sigma }^{c}Tr\,\left( T^{a}\left\{ T^{b}T^{c}\right\} \right) .
\end{equation}

There is also one gravitational term that comes from $e_{4}\left( P\right) $
\begin{equation}
\epsilon ^{\mu \nu \rho \sigma }\theta ^{a}Tr\,\left( T^{a}\right) R_{\mu
\nu }^{cd}R_{\rho \sigma cd} .
\end{equation}
Therefore gravitational anomalies are absent if one further imposes the
condition \cite{A-W}
\begin{equation}
Tr\left( T^{a}\right) =0 .
\end{equation}

One can make similar analysis for anomalies on higher dimensional manifolds.
It can be immediately seen that in ten dimensions the first non-vanishing
trace in the heat-kernel expansion would result from a term of the form
\begin{equation}
Tr\left( \gamma _{11}\gamma ^{\mu _{1}\mu _{2}}\ldots \gamma ^{\mu
_{9}\mu_{10}} T^{a_{1}}\cdots T^{a_{5}}\right) F_{\mu _{1}\mu
_{2}}^{a_1}\cdots F_{\mu_{9}\mu_{10}}^{a_5} ,
\end{equation}
which is obviously related to the well known hexagon diagram \cite{A-W} .

We therefore reach the conclusion that chiral gauge anomalies arise in the
path integral formulation because the fermionic measure is not invariant
under chiral rotations \cite{Fuji, Bala, Alv, A-W} , while in the spectral
action they arise because of the non-invariance of the trace operator. The
anomaly cancellation conditions are, however, identical in both cases.

\section{Acknowledgments}

I would like to thank Alain Connes, J\"urg Fr\"ohlich, Jose Garcia-Bondia
and Thomas Sh\"ucker for useful discussions. I would also like to thank
Daniel Kastler for inviting me to contribute to this volume.

\end{document}